\documentclass[showpacs,preprint,preprintnumbers,amsmath,amssymb,floatfix]{revtex4-1}
\usepackage{graphicx}
\usepackage{amssymb}
\usepackage{bm}
\usepackage{color}
\begin{document}

\title{55~Tesla coercive magnetic field in frustrated Sr$_3$NiIrO$_6$}

\author{John~Singleton$^{1}$, Jae~Wook~Kim$^{1}$, Craig~V.~Topping$^{1,2,3}$, Anders Hansen$^1$, Eun-Deok~Mun$^{1,4}$, Saman~Ghannadzadeh$^{3}$, Paul~Goddard$^{5}$, Xuan Luo$^{6,7}$, Yoon~Seok~Oh$^{6}$, Sang-Wook~Cheong$^{6}$, Vivien~S.~Zapf$^{1}$}

\address{$^{1}$National High Magnetic Field Laboratory (NHMFL), MS~E536, Los Alamos National Laboratory, Los Alamos, NM 87545, USA}
\address{$^{2}$Department of Chemistry, University of Edinburgh, Edinburgh, Midlothian EH8~9YL, United Kingdom}
\address{$^{3}$University of Oxford, Department of Physics, The Clarendon Laboratory, Parks Road, Oxford, OX1~3PU, United Kingdom}
\address{$^4$Department of Physics, Simon Fraser University, Burnaby, BC, V5A~1S6, Canada}
\address{$^{5}$Department of Physics, University of Warwick, Gibbet Hill Road, Coventry, CV4~7AL, United Kingdom}
\address{$^6$RCEM \& Dept. of Physics and Astronomy, Rutgers University, Piscataway, NJ~08854, USA}
\address{$^7$POSTECH, Pohang University of Science and Technology, San 31 Hyoja-dong, Nam-gu, Pohang-si, Gyungbuk, 790-784, Republic of Korea}
\pacs{71.70.Ej, 75.30.Gw, 75.50.Vv}

\begin{abstract}
We have measured extremely large coercive magnetic fields
of up to 55~T in Sr$_3$NiIrO$_6$, with a switched magnetic moment 
$\approx 0.8~\mu_{\rm B}$ per formula unit.
As far as we are aware, this is the largest coercive field observed thus far.
This extraordinarily hard magnetism has a completely different origin from
that found in conventional ferromagnets.
Instead, it is due to the evolution of a frustrated antiferromagnetic state
in the presence of strong magnetocrystalline anisotropy due to the 
overlap of spatially-extended Ir$^{4+}$ 5$d$ orbitals with 
oxygen 2$p$ and Ni$^{2+}$ 3$d$ orbitals.
This work highlights the unusual physics that can result from combining 
the extended $5d$ orbitals in Ir$^{4+}$ with the frustrated behaviour of triangular lattice antiferromagnets.
\end{abstract}

\maketitle
\section{Introduction}
Oxides containing iridium in the Ir$^{4+}$ ($5d^5$) state have received much recent 
attention because the energy scales for spin-orbit interactions (SOIs), Coulomb repulsion, 
and crystalline-electric fields are all very 
similar~\cite{kim09, lovesy12, chapon11,Shitade09,Laguna10,Chikara09,Chaloupka10,Wan11,Choi12,kim08,Kim13}.
This unusual situation, in comparison to analogous $3d$ systems,
results from a decrease in the strength of correlation effects
and an increase in SOIs as one descends the periodic table.
Usually, the Coulomb repulsion and SOIs are responsible for Hund's rules that 
determine the groundstates of magnetic ions; 
however, in these $5d$ (and some $4d$) systems, the 
competition between the three similarly-sized 
energy scales can result in exotic magnetic 
states~\cite{kim09,lovesy12,chapon11,Shitade09,Laguna10,Chikara09,Chaloupka10,Wan11,Choi12,kim08,Kim13}, 
leading to proposals for the use of Ir$^{4+}$-based systems 
in quantum information processing and other novel 
applications (see Refs.~\cite{lovesy12,Choi12} and references therein). 
In this paper, we report another manifestation of unusual $5d$ physics:
extremely large coercive magnetic fields 
of up to 55~T in Sr$_3$NiIrO$_6$, with a switched magnetic 
moment of about $0.8 \mu_{\rm B}$ per formula unit.
As far as we are aware, this is the largest coercive field measured thus far,
and a factor $\sim 5$ times those reported recently in novel, ultra-hard ferromagnets~\cite{jesche14}.
This exemplary hard magnetism is, however, very different
from that found in conventional ferromagnets in that it results from
an unusual, frustrated, antiferromagnetic ground state
that incorporates a relatively large magnetocrystalline anisotropy.

\begin{figure}
\centering
\includegraphics[width=0.69\textwidth]{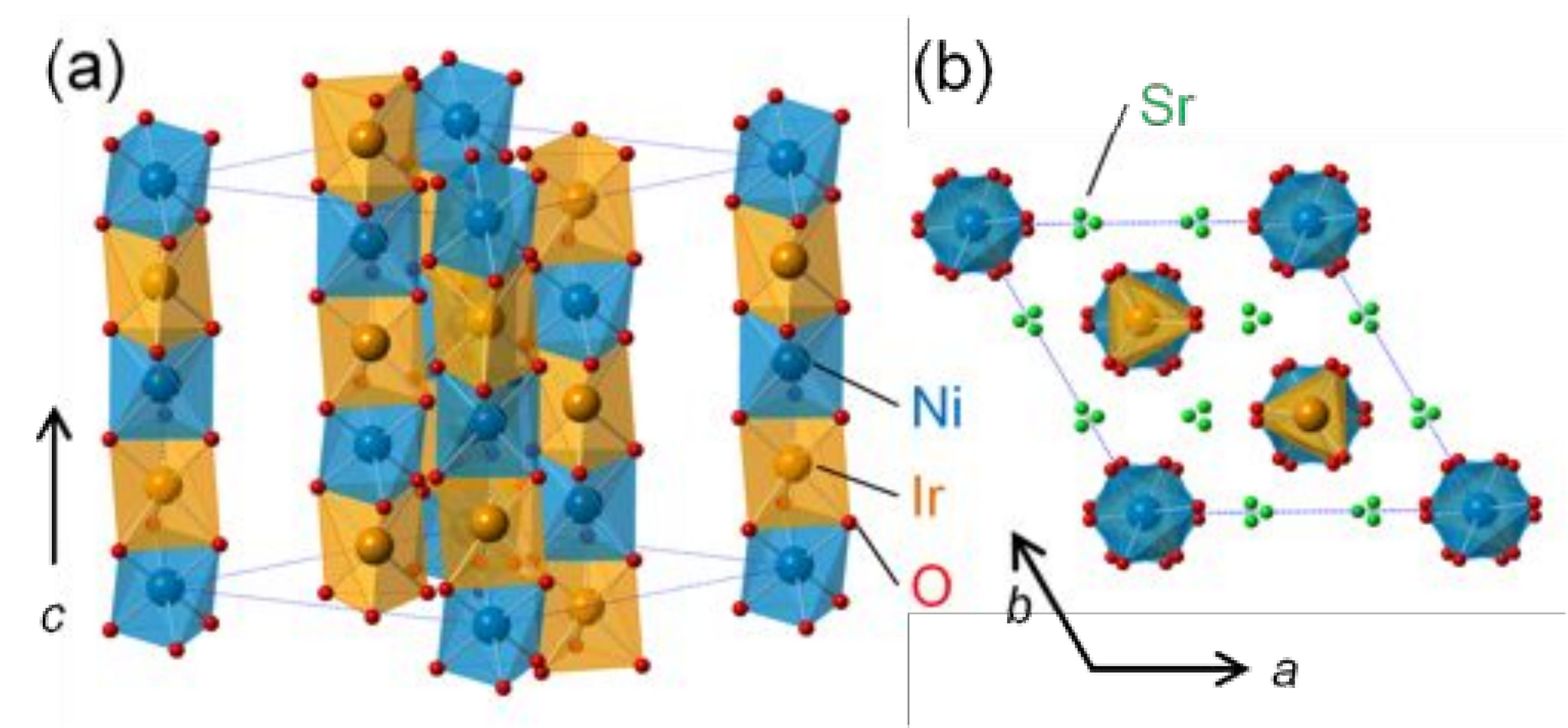}
\includegraphics[width=0.30\textwidth]{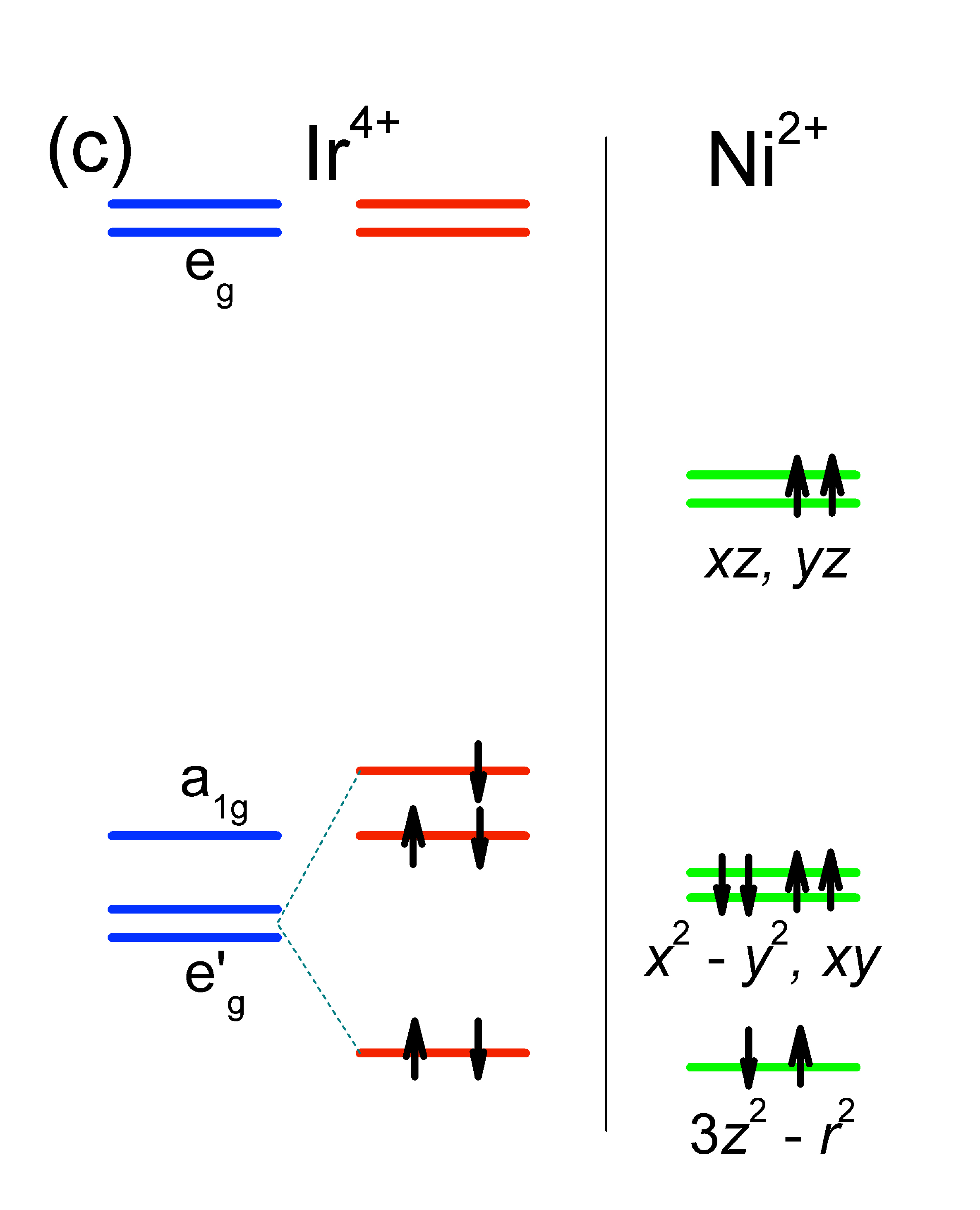}
\caption{{\bf Structure and energy levels.} The crystal structure of Sr$_3$NiIrO$_6$ 
as viewed from the (a)~$[110]$ and (b)~$[001]$ directions.
The Ni$^{2+}$ and Ir$^{4+}$ ions are within oxygen trigonal bipyramids 
(blue) and oxygen octahedra (orange), respectively.
For clarity, Sr ions are not shown in (a).
(c)~Schematic level diagrams for Ir$^{4+}$ and Ni$^{2+}$ (based on
Refs~\cite{Zhang10,Sarkar10,Wu13}). Unperturbed 
Ir$^{4+}$ levels are shown in blue; a combination of antiferromagnetic
interactions and spin-orbit coupling splits the e$^{\prime}_{g}$
doublet to give the configuration shown in red.
Electron spins are shown as arrows:
the effective spin of Ni$^{2+}$ is $S=1$
and that of Ir$^{4+}$ is $S=\frac{1}{2}$
in the opposite direction.}
\label{Structure}
\end{figure}

The salient structural
details~\cite{Nguyen95} are shown in Fig.~\ref{Structure}.
The Ni$^{2+}$ and Ir$^{4+}$ 
magnetic ions in Sr$_3$NiIrO$_6$ occupy oxygen cages that alternate in 
chains parallel to the $c$-axis (Fig.~\ref{Structure}(a)). 
These chains are in turn arranged in a hexagonal pattern 
in the $ab$-plane (Fig.~\ref{Structure}(b)) \cite{Nguyen95}.
The Ni$^{2+}$ is surrounded by a trigonal bipyramid of 
oxygen atoms, while the Ir$^{4+}$ ion sits in an octahedral oxygen cage.
Magnetic frustration is intrinsic to this structure, 
and can result from antiferromagnetic interactions 
within the triangular lattice in the $ab$-plane, 
and from frustration between nearest-neighbour and 
next-nearest-neighbour interactions along the $c$-axis chains.
Electronic structure calculations have suggested both 
possibilities for Sr$_3$NiIrO$_6$~\cite{Zhang10,Sarkar10,Wu13};
the same calculations indicate that the overlap of 
spatially-extended Ir$^{4+}$ $5d$ orbitals with 
oxygen $2p$ and Ni$^{2+}$ $3d$ orbitals leads to a 
magnetocrystalline anisotropy energy~$\approx 13.5$~meV.
Schematic energy level diagrams for Ir$^{4+}$ and Ni$^{2+}$ are shown in Fig.~\ref{Structure}(c);
the effective spin of Ni$^{2+}$ is $S=1$
and that of Ir$^{4+}$ is $S=\frac{1}{2}$~\cite{Zhang10,Sarkar10,Wu13}.

\begin{figure}[htbp]
\centering
\includegraphics[width=0.375\textwidth]{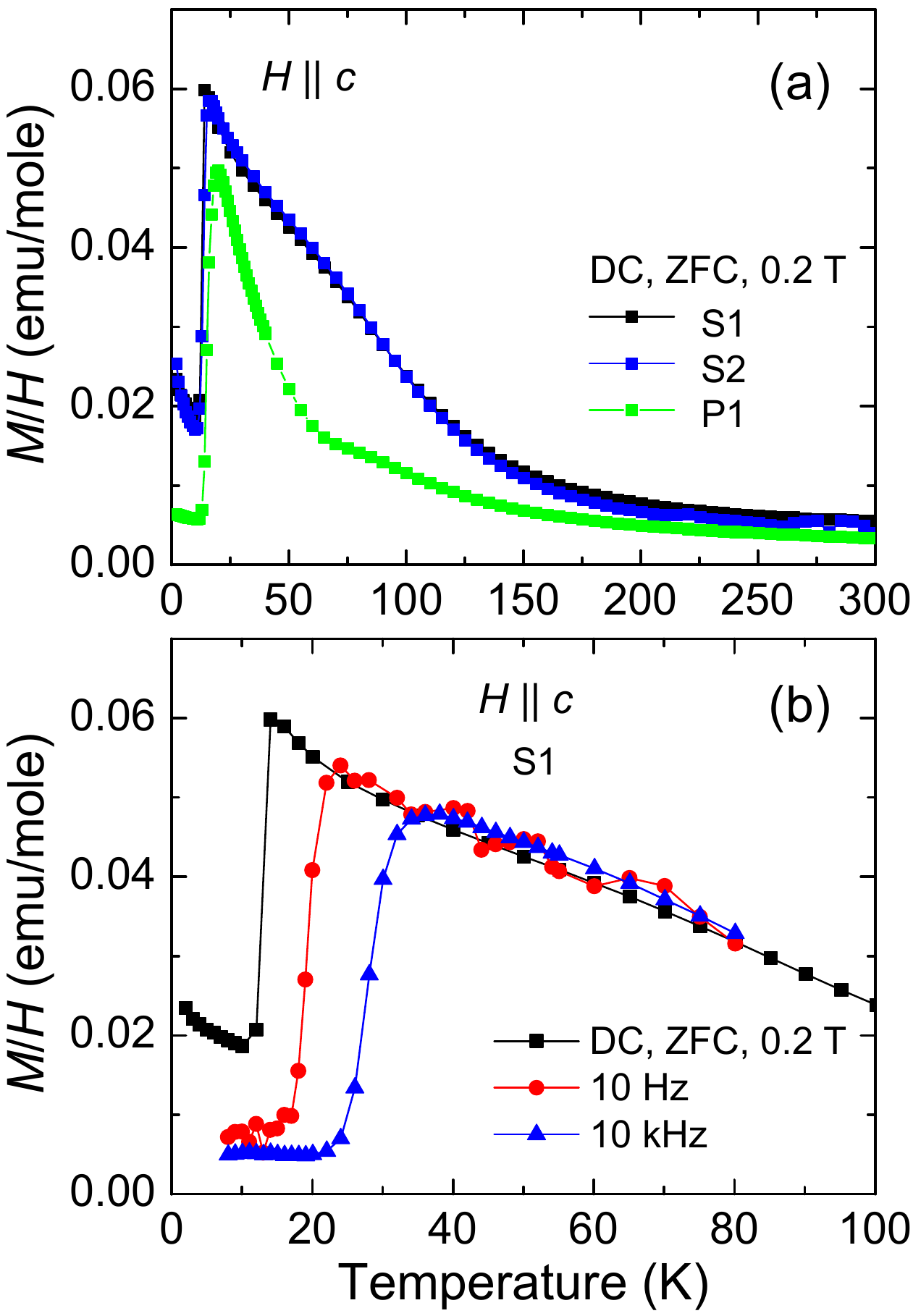}
\includegraphics[width=0.615\textwidth]{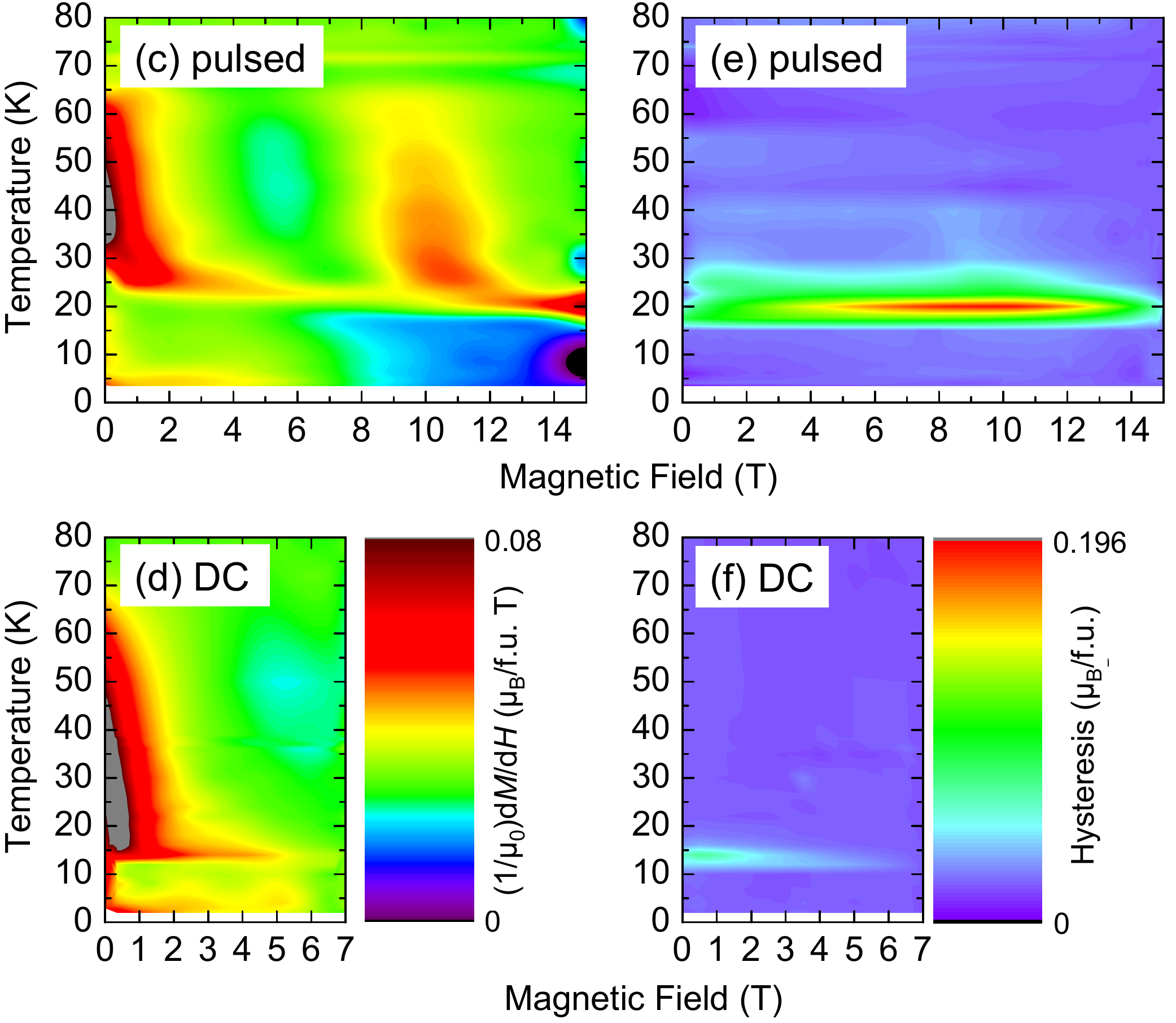}
\caption{{\bf Frustration-induced slow dynamics.} 
(a)~DC magnetic susceptibility ($\chi = M/H$) as a function of temperature $T$ for Sr$_3$NiIrO$_6$
single-crystal samples S1 and S2 and polycrystalline sample P1.
The onset of magnetic order is marked by a broad hump
superimposed on the background slope and centered on $T \approx 85$~K.
(b)~$T$-dependent $\chi$ data
for Sr$_3$NiIrO$_6$ single crystal sample S1
measured DC (black) and using an AC susceptometer
operated at frequencies of 10~Hz (red) and 10~kHz (blue).
The sharp fall in $\chi$ shifts to higher $T$
as the frequency increases.
The contour plots show ${\rm d}M/{\rm d}H$ versus 
$\mu_0 H$ and $T$ for polycrystalline sample P2 measured 
(c)~in a pulsed magnetic field with a 
sweep rate $\approx 3500$~T/s and (d)~in a SQUID (DC) magnetometer 
with a magnetic field 
sweep rate of 0.008~T/s.
Corresponding plots of the hysteresis between
falling and rising magnetic field sweeps, defined as 
hysteresis~$=\left[M(H,T)_{\rm falling} - M(H,T)_{\rm rising}\right]$, are shown in (e)~(pulsed field) and 
(f)~(SQUID). Note that as the field sweep rate increases, features 
in ${\rm d}M/{\rm d}H$ and regions of pronounced 
hysteresis are pushed to higher $T$. }
\label{MT}%
\end{figure}%
\section{Results}
\subsection{Low-field magnetization and dynamics}
Our study involves 10 different samples
of Sr$_3$NiIrO$_6$ from several different batches
encompassing minor variations in growth conditions. The results
for all samples are rather consistent; typical data from
four single crystals (labelled S1 to S4) and three polycrystalline
samples (P1, P2, P3) are shown in the figures that follow.
Note that sample S2 is a very small crystal with exceptionally
clean hexagonal faces containing virtually no visible defects; as will be seen below,
this apparent cleanliness
is of relevance to the timescales over which the remanent magnetization persists. 

Typical examples of low-magnetic-field DC magnetization data 
for Sr$_3$NiIrO$_6$ (samples S1, S2 and P1) are shown in Fig.~\ref{MT}. 
Magnetic order sets in below 85~K, 
being marked by a broad feature in magnetization ($M$) vs. 
temperature ($T$) data (Fig.~\ref{MT}(a)), and additional peaks in 
elastic neutron scattering~\cite{Flahaut03,Mikhailova12,LeFrancois14}.
The neutron diffraction data are consistent with either a partially-disordered 
antiferromagnet~\cite{Mekata77} or an amplitude-modulated 
antiferromagnetic state at low $T$.
Despite the onset of long-range order at 85~K, the heat capacity 
is nearly featureless between 4 and 250~K \cite{Mikhailova12}.

Fig.~\ref{MT}(b) shows the DC magnetic susceptibility
of Sr$_3$NiIrO$_6$ single crystal sample S1, in 
an applied field of $\mu_0 H =0.2$~T, and the AC susceptibility for 
an oscillating field amplitude of $10^{-3}$~T at 10~Hz and 10~kHz.
In all three data sets, the sharp drop in $M$ already seen in
Fig.~\ref{MT}(a) is observed as $T$ is lowered.
However, the temperature at which the drop occurs is
frequency dependent; the half-height point
occurs at 28~K for 10~kHz, 20~K 
for 10~Hz, and 13~K for DC measurements.
Further evidence of slow dynamics over a similar
temperature range is seen in the contour plot 
in Figs.~\ref{MT} (c) to (f). Here we show ${\rm d} M(H,T)/{\rm d}H$
on the initial field upsweep following zero-field cooling for
polycrystalline sample P2.
These data were measured in a 7~T SQUID magnetometer
with a field-sweep rate $\approx 0.008$~T/s (Fig.~\ref{MT}(d)) and in a 
pulsed magnetic field with a sweep rate $\approx 3500$~T/s 
(Fig.~\ref{MT}(c)), where the pulse shape is shown in Fig.~\ref{fieldbolg}(a).
Corresponding plots of the hysteresis in $M$ between rising and
falling magnetic field sweeps are shown in (e)~(pulsed field) and 
(f)~(SQUID). As the field sweep rate increases
(and therefore the timescale of the measurement decreases), features 
in ${\rm d}M/{\rm d}H$ and regions of pronounced 
hysteresis are pushed to higher $T$.
Thus, this dependence of $M$ on sweep rate is analogous to the
frequency dependence seen in the AC susceptibility data (Fig.~\ref{MT}(b)),
where decreasing the measurement timescale 
forces features in $M$ to higher $T$.
Similar phenomena have been seen in previous
experiments within the ordered phase; 
as $T$ was lowered below 30~K, a frequency-dependent 
crossover occurred to hysteretic magnetic behaviour as a 
function of $T$ and $H$, as well as a strong 
frequency dependence of the AC susceptibility~\cite{Flahaut03,Mikhailova12,LeFrancois14}.
A number of other frustrated antiferromagnets isostructural with Sr$_3$NiIrO$_6$ show 
qualitatively similar frequency-dependent magnetic 
behaviour~\cite{Kageyama98,Kawasaki99,Niitaka01,Stitzer02,Rayaprol03,Loewenhaupt03,Hardy03,Hardy04,Mohapatra07,Choi08,Agrestini08,Agrestini11,Kim14}.

\begin{figure}
\centering
\includegraphics[width=.95\textwidth]{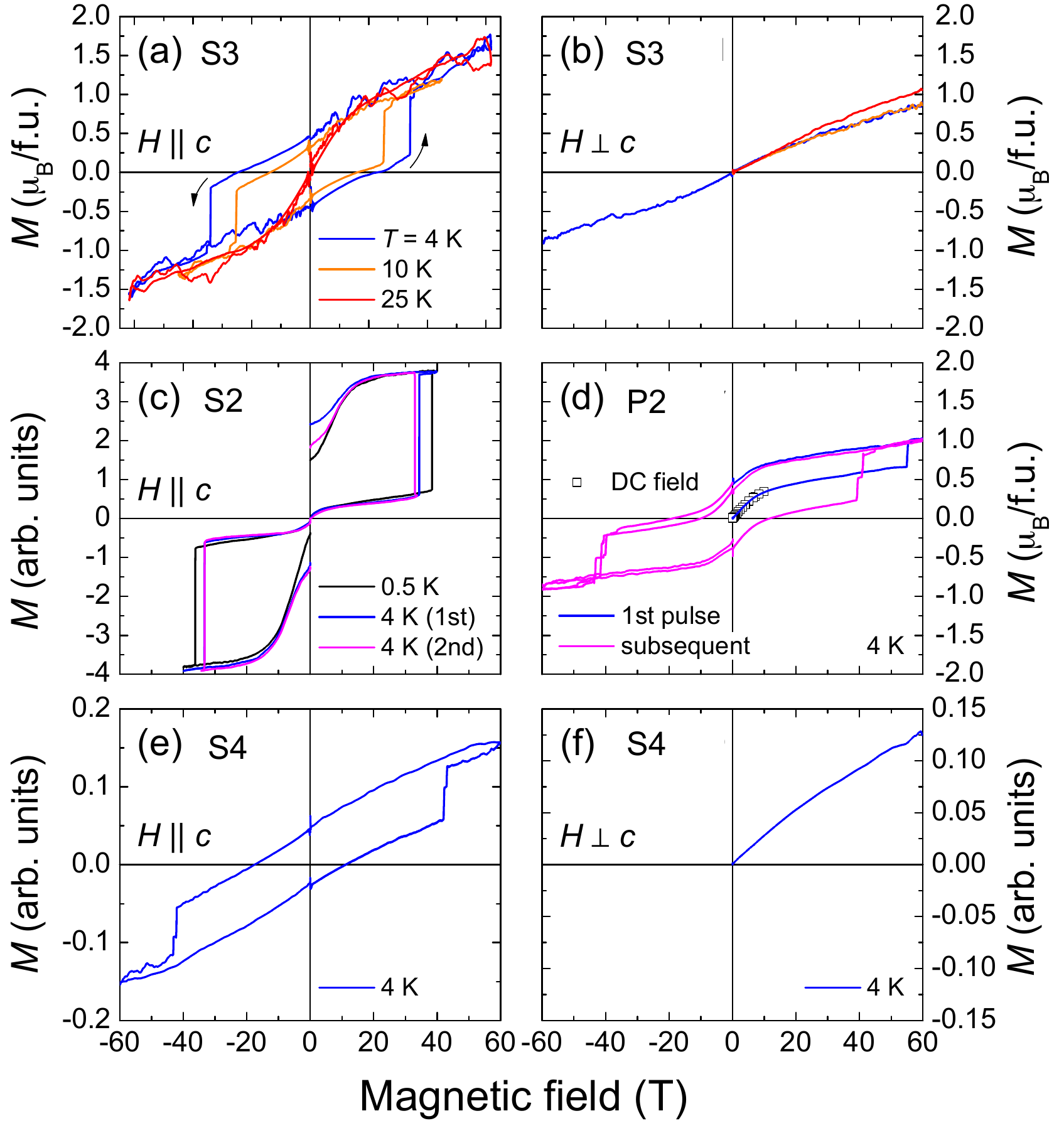}
\caption{{\bf Hysteresis loops and large coercive fields.} 
Magnetization $M$ as a function of magnetic field $\mu_0 H$ 
measured in a series of pulses using a capacitor-bank-driven 65~T pulsed magnet.
Sample numbers, field directions 
and measurement temperatures $T$ are given in each section of the figure;
the vertical jumps in $M$ occur at the coercive field, $\mu_0 H_{\rm c}$.
In (b), (f) no hysteresis is observed in the ${\bf H} \perp {\bf c}$ configurations.
In (a), (b), and (d), absolute magnetization values, measured in
commercial SQUID and vibrating-sample magnetometers, 
were used to calibrate the pulsed field data.}
\label{Hysteresis}
\end{figure}
\subsection{Large coercive fields}
Large coercive fields are demonstrated in
Fig.~\ref{Hysteresis}, which shows pulsed-field
$M(H)$ data for four representative 
samples of Sr$_3$NiIrO$_6$ (the field-pulse profile
is shown in Fig.~\ref{fieldbolg}(a)).
Data for Figs.~\ref{Hysteresis}(a), (b), and (d) are calibrated 
against absolute measurements on the same samples in quasistatic 
magnetic fields of up to 13~T.

\begin{figure}
\centering
\includegraphics[width=0.75\textwidth]{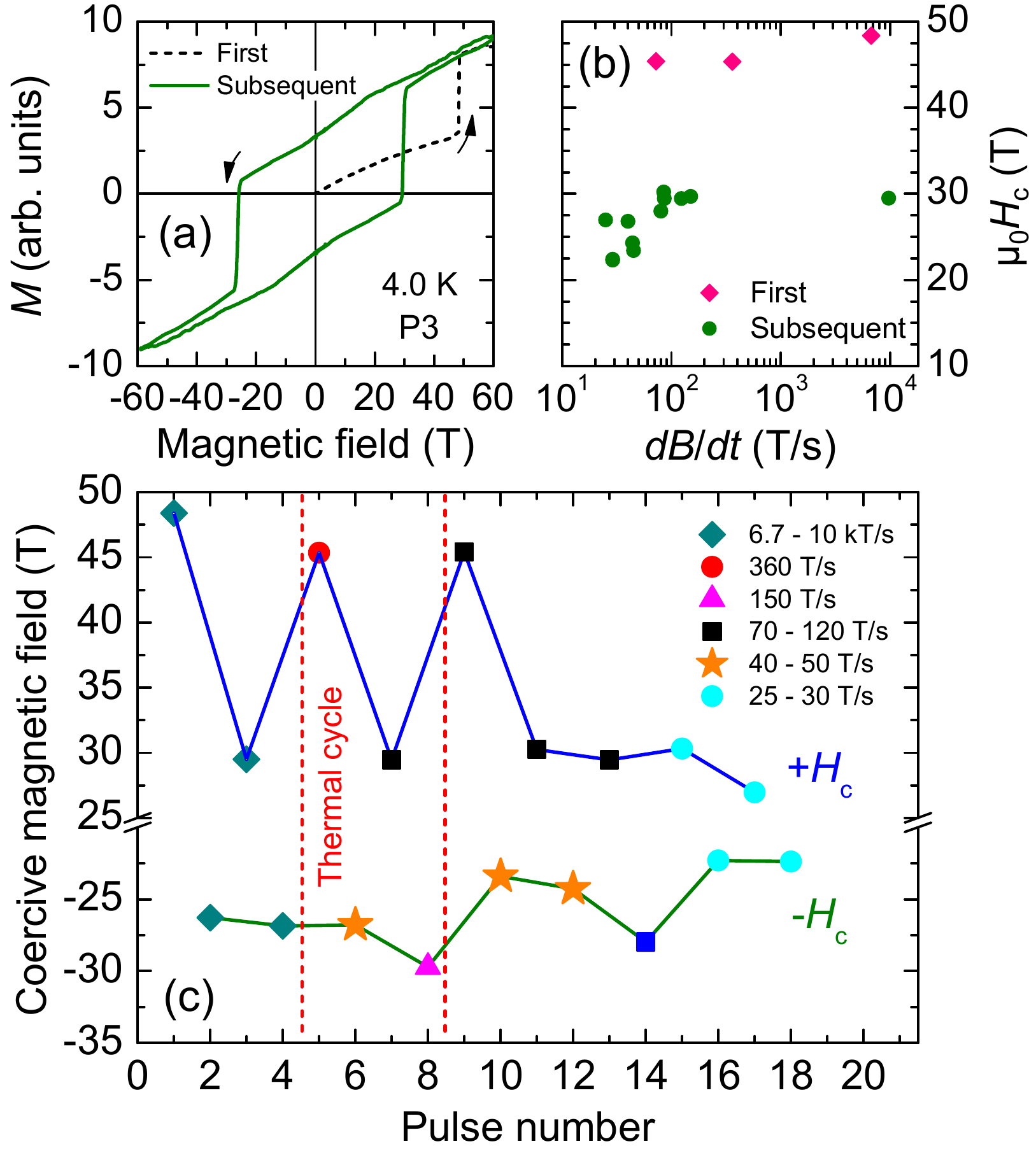}
\caption{{\bf Sweep-rate dependence of the coercive field.} 
(a) $M(H)$ data for polycrystalline sample P3 measured in a capacitor bank-driven pulsed magnet.
Black dotted lines denote the initial pulse after zero-field cooling whilst the green 
solid line indicates data for subsequent pulses at 4~K.
(b) Sweep-rate dependence of coercive magnetic field ($\mu_0 H_{\rm c}$).
(c) History dependence of coercive magnetic field ($\mu_0 H_{\rm c}$) under 
various sweep rates (plotted as a function of pulse number).
For sweep rates larger than 360~T/s, a capacitor-bank-driven pulsed
magnet was used; other data were taken using the generator-driven
60~T Long-Pulse Magnet (see Fig.~\ref{fieldbolg} for magnetic field profiles).
Twice during this experiment, the sample was warmed to room temperature 
(denoted by red vertical dashed lines).}
\label{field}
\end{figure}

Though the behaviour of all samples is similar,
showing that the large coercive fields are intrinsic to Sr$_3$NiIrO$_6$,
there are some detail differences from sample to sample.
In Fig.~\ref{Hysteresis}(a) (single crystal S3), the magnetic hysteresis for 
${\bf H} \parallel {\bf c}$ is superimposed on a linear $M(H)$ background, 
and shows a large $\mu_0 H_{\rm c} = 34.1$~T and a remanent 
moment of $0.45 \mu_{\rm B}$ per formula unit at zero field. 
The magnetization jump at 34.1~T is $\approx 0.8\mu_{\rm B}$ per formula unit.
The data in Fig.~\ref{Hysteresis}(e) (Sample S4)
show an even larger $\mu_0 H_{\rm c} = 42.7$~T.
As mentioned above, sample S2 is a very small single crystal with exceptionally clean faces,
and Fig.~\ref{Hysteresis}(c)) shows that it exhibits qualitatively different behaviour: the coercive 
field is very large (38.4~T), as in other samples, but $M$ relaxes 
during the millisecond timescales (see Fig.~\ref{fieldbolg}(a))
of the downsweep of the magnetic field.
All subsequent field pulses after the first one show jumps 
of similar height, consistent with relaxation of $M$ 
back to zero between field pulses. 
On the other hand, in Fig.~\ref{Hysteresis}(a), (b), (d), (e), and (f) 
the first upsweeps following a zero-field 
cool result in an $M$ jump about half the height of subsequent jumps, 
showing that the latter represent a complete reversal of $M$.
For those samples, two consecutive magnetic fields 
sweeps in the same direction produce no jump (not shown).
In polycrystalline sample P2 (Fig.~\ref{Hysteresis}(d)), the $M(H)$ data 
also show some curvature at low fields, but $M$ does not relax 
back to zero between field pulses. (As will be shown in
Fig.~\ref{field}(a) below, polycrystalline sample
P3 shows similar behaviour, but superimposed on a much less
curved background.)
In the initial up-sweep, $M$ jumps at a record-high 
magnetic coercive field of $\mu_0H_{\rm c}=55$~T after zero-field 
cooling; in subsequent pulses, the $M$ jump doubles in height, 
but is observed at lower fields (40~T).
Note that all samples exhibit a higher $H_{\rm c}$ value in the initial field sweep after zero-field cooling,
compared to those seen in subsequent field sweeps; we shall return to this behaviour below. 
Finally, we remark that the jumps in $M$ are sharp, occurring over a small range of field 
($< 1$~T at 4~K). In particular, the transition width in Fig.~\ref{Hysteresis}(c) is $\lesssim 0.03$~T.

There is no systematic dependence of the magnetic 
hysteresis on details of the growth method in the samples 
shown in Fig.~\ref{Hysteresis} or in the three other samples measured,
suggesting that all samples have the same Ni/Ir ratio. 
There does, however, appear to be a pinning effect that 
stabilizes $M$ in most of the samples;
in such cases, the remanent magnetization was found to persist unchanged
for at least 24 hours at $T=4$~K.
By contrast, sample S2 in 
Fig.~\ref{Hysteresis}(c) shows magnetic relaxation on millisecond timescales
({\it i.e.} similar timescales to the downsweep of the pulsed field; see Fig.~\ref{fieldbolg}(a)).
In addition, none of the samples studied 
appear to show signs of saturation of the 
magnetization even in fields of 65~T.
The anisotropy of the hysteresis is illustrated in Figs.~\ref{Hysteresis}(b) and (f); here 
$M(H)$ of single crystals S3 and S4 is shown for ${\bf H}\perp {\bf c}$.
For this field direction, no jump in $M$ or hysteresis is observed.
In both single crystals, $M$ is comparable in size at 
65~T for ${\bf H} \perp {\bf c}$ and ${\bf H} \parallel {\bf c}$.
\subsection{Dynamics, and history- and temperature dependences of the coercive field}
In view of the field-sweep-rate dependence of the low-field magnetization
noted earlier (Fig.~\ref{MT}(b)-(f)), it is interesting to
see if the hysteresis loops are affected by similar dynamical issues, and so
Fig.~\ref{field} features $M(H)$ measurements of polycrystalline sample P3 
in the Los Alamos generator-driven 60~T Long-Pulse Magnet, which 
provides sweep rates between 25 and 360~T/s, and in a capacitor-driven 65~T
magnet with faster sweep rates (pulse shapes are shown in Fig.~\ref{fieldbolg}).
Fig.~\ref{field}(a) once again demonstrates that the coercive field is higher on the initial
field sweep after zero-field cooling (c.f. Figs.~\ref{Hysteresis}(c) and (d)), 
whilst (b) shows that this trend also occurs at lower field-sweep rates.
The history dependence of the
coercive field throughout a sequence of field pulses of varying sweep rate
and a couple of thermal cycles to room temperature and back
is illustrated in Fig.~\ref{field}(c).
All of these data show that the magnetic hysteresis 
loop is robust at least down to 25~T/s and illustrate that 
on magnet pulses subsequent to the initial pulse 
following zero-field cooling, $H_{\rm c}$ is slightly 
smaller in lower field sweep rates; this was observed in all measured samples.

\begin{figure}
\centering
\includegraphics[width=0.495\textwidth]{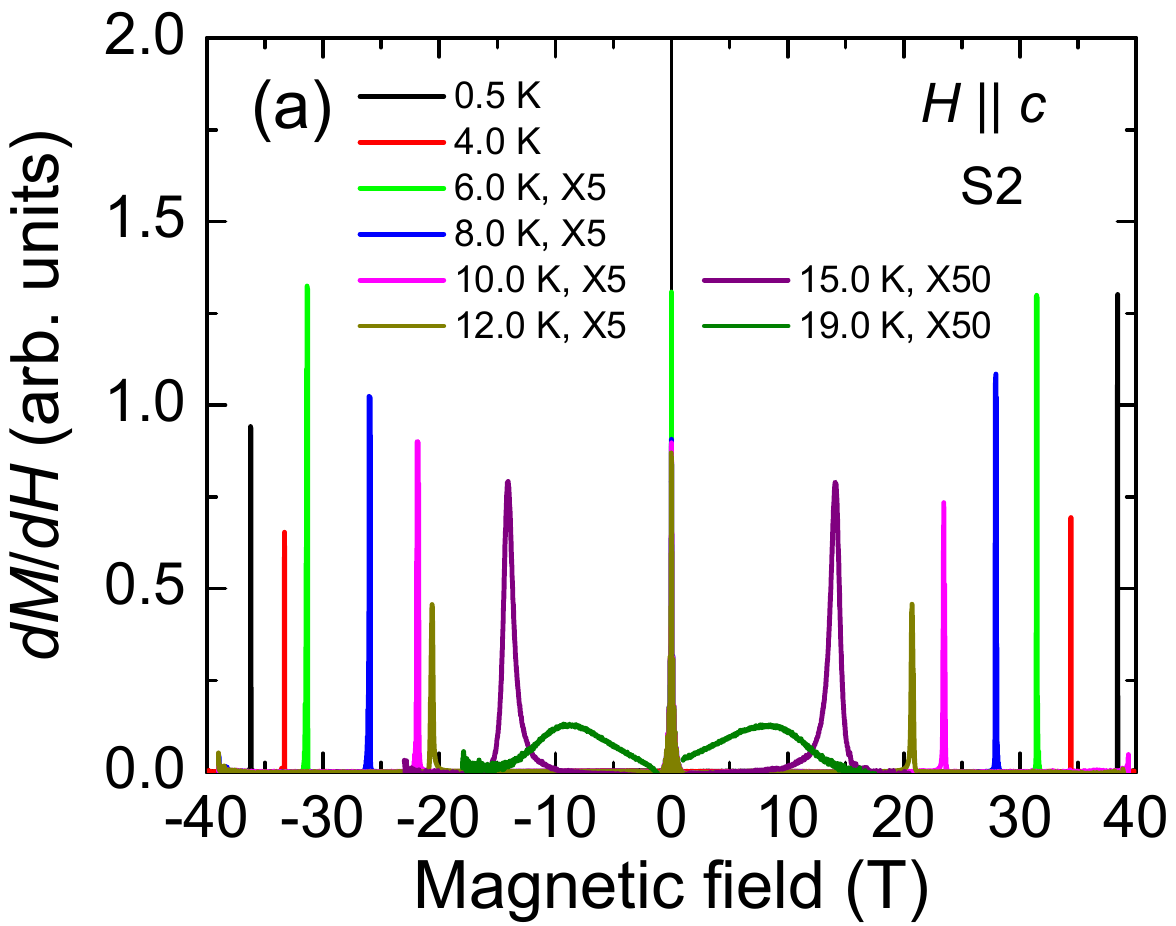}
\includegraphics[width=0.495\textwidth]{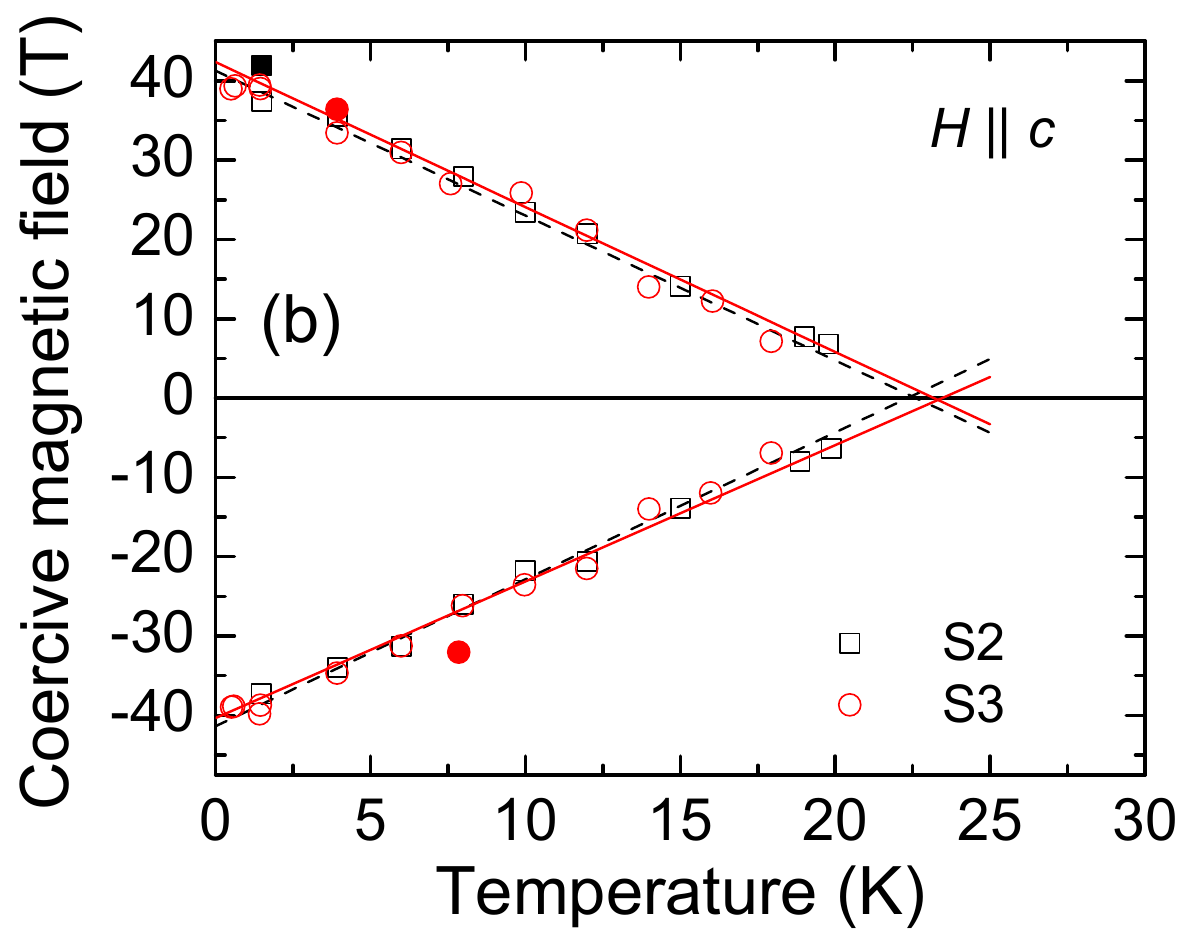}
\caption{{\bf Delineating the region of large coercive field.}
(a) ${\rm d}M/{\rm d}H$ for single-crystal sample S2
at several $T$ for ${\bf H}\parallel {\bf c}$ (c.f. Fig.~\ref{Hysteresis}(a,c)).
The peaks denote the sharp change in $M$ at the coercive field;
note that multiplication factors of 5 and 50 
are applied for 6, 8, 10~K and 15, 19~K data, respectively as the peak broadens with increasing $T$.
(b) $T$-dependence of the coercive field for single crystals S2 and S3.
Filled symbols indicate the result taken in the initial pulse after zero-field cooling from room $T$; 
these show slightly larger $H_{\rm c}$ values compared those measured in subsequent pulses
(see Fig.~\ref{field}).
Lines are linear fits to the data.}
\label{PhaseDiagram}
\end{figure}

The $T$-dependences of $M(H)$ for ${\bf H} \parallel {\bf c}$ 
and ${\bf H} \perp {\bf c}$ are shown in Figs.~\ref{Hysteresis}(a) 
and (b) for single crystal S3.
For ${\bf H}\perp {\bf c}$, $M$ for $T < 25$~K is suppressed 
compared to the value at 25~K, consistent with 
antiferromagnetic exchange interactions~\cite{Cullity}.
For ${\bf H}\parallel {\bf c}$, hysteresis is observed at 10~K but has disappeared by 25~K.
The $T$-dependence of $H_{\rm c}$ is given in 
more detail in Fig.~\ref{PhaseDiagram}(a), which shows 
${\rm d}M/{\rm d}H$ data for ${\bf H} \parallel {\bf c}$ (sample S2); 
peaks occur at the coercive field, $H_{\rm c}$.
Fig.~\ref{PhaseDiagram}(b) shows that $H_{\rm c}$ 
decreases linearly with increasing $T$ (data from samples S2 and S3), 
extrapolating to zero at $\approx 25$~K. 
This is close to the onset of hysteretic behaviour 
and the frequency-dependent fall in $M$ (see Fig.~\ref{MT}(b-f)), linking
the observation of finite coercive field 
with the low-temperature phase of the antiferromagnetic groundstate.
The almost identical variations of $H_{\rm c}$ with temperature
in the two samples show that the value of this field
and its temperature dependence are intrinsic to Sr$_3$NiIrO$_6$ single crystals,
whereas the persistence (or not) of the remanent magnetization depends on extrinsic
factors such as defects or disorder (the same samples
are used in Figs.~\ref{Hysteresis}(a),(c)).

Finally, we note that a coercive field $\mu_0 H_{\rm c} \approx 22$~T
was measured using polycrystalline samples of Sr$_3$NiIrO$_6$ 
in Ref.~\cite{Flahaut03}. All of our polycrystalline samples
exhibit low-temperature coercive fields  $\mu_0 H_{\rm c} \approx 40$~T,
very similar to those seen in the single crystals (see Figs.~\ref{Hysteresis}
and \ref{field}). Therefore, the difference between
polycrystalline and single-crystal samples cannot account for
the much lower values of $H_{\rm c}$ observed
in Ref.~\cite{Flahaut03} compared to those in this work.
\section{Discussion}
\subsection{Summary of results}
Combining the $H$- and $T$-dependent magnetization 
studies of Sr$_3$NiIrO$_6$, 
a picture emerges of an antiferromagnetic order~\cite{LeFrancois14} 
for $T \lesssim 30$~K and 
$H=0$ that evolves in applied magnetic fields.
This evolution is manifested most
markedly in the sharp jump in magnetization that occurs on rising fields. 
The field position $\mu_0H_{\rm c}$ of the jump shows a 
similar history and field-sweep-rate
dependence for all samples (Figs.~\ref{Hysteresis}, \ref{field}, \ref{PhaseDiagram}), 
with slightly smaller values of $H_{\rm c}$
occurring at lower sweep rates. 
This strongly suggests that the magnetization jump at $H_{\rm c}$ 
is an intrinsic property of Sr$_3$NiIrO$_6$, with the sweep-rate dependence
being caused by sluggish kinetics associated with 
the magnetic frustration intrinsic to this structural family~\cite{Kim13};
the same frustration is responsible for the slow magnetic relaxation (Figs.~\ref{MT}(c-f)) and the strong variation 
($13 - 28$~K) of the fall in $M$ with frequency (Fig.~\ref{MT}(b)).
By contrast, in systems such as (Sm,Sr)MnO$_3$~\cite{Fisher04},
where the magnetization jumps are not intrinsic, 
but associated with quenched disorder,
the smaller the field-sweep rate, the {\it larger} the field needed to
realize the transition, the opposite of what we observe in Sr$_3$NiIrO$_6$ (Fig.~\ref{field}). 

It is likely that the millisecond-timescale magnetic relaxation observed in the 
hysteresis loops for sample S2 (Fig.~\ref{Hysteresis}(c)) is also intrinsic 
behaviour due to frustration, whilst samples S3, S4, P2 (Figs.~\ref{Hysteresis}(a), (d), (e))
and P3 (Fig.~\ref{field}(a)) and all other samples studied
exhibit pinning or freezing of the magnetization or 
an order-by-disorder mechanism, as seen at lower fields in isostructural 
Ca$_3$CoMnO$_6$~\cite{Kiryukhin09}.
In the latter mechanism, magnetic-site disorder disrupts longer-range
interactions, leading to a reduction of frustration~\cite{Kiryukhin09}.
\subsection{Comparison with ferromagnets and other systems}
Magnets with high coercivity, {\it i.e.}, ``hard" magnets, 
are important for a wide range of applications involving magnetic 
actuation and induction, such as loud speakers, wind 
turbines and electric motors~\cite{Cullity}. 
In rare-earth iron boride magnets, among the 
hardest commercial magnets, $\mu_{0}H_{\rm c}$ can be as large as 1.5~T at room temperature~\cite{Coey11}.
At cryogenic temperatures, magnetic hysteresis effects can 
extend to higher fields, {\it e.g.}, up to 10~T 
in the colossal magnetoresistance manganites, and 
in Li$_2$(Li$_{1-x}$Fe$_x$)N, Gd$_5$Ge$_4$, Ga-doped CeFe$_2$, LuFe$_2$O$_4$, 
and Fe$_{1/4}$TaSe$_2$~\cite{jesche14,Hardy04,Autret03,Morosan07,Haldar08,Wu08,Ko11}.
Large coercive fields in such magnets are typically caused by magnetocrystalline
anisotropy due to SOI~\cite{Coey11},
whilst magnetic hysteresis results from a combination of
domain dynamics and anisotropy~\cite{Cullity}. 
In traditional ferromagnets, domains result from competition between 
the short-range exchange interactions that prefer parallel spin
alignment, and the free-energy penalty of maintaining a magnetic field 
in an extended region of space around the sample~\cite{Cullity}.
The effect of this competition is that the energy scale 
for switching magnetic domains can be orders of magnitude 
smaller than those of the nearest-neighbor 
ferromagnetic exchange interactions. 
Generally, the microscopic order in a traditional 
ferromagnet does not change significantly around the hysteresis loop as the domains 
change direction and/or the domain walls move~\cite{Cullity}.

By contrast, the conventional phenomenology of ferromagnetic domains is
{\it not} involved in the notable $H_{\rm c}$ values observed for
Sr$_3$NiIrO$_3$ ($35-55$~T, this work) and in the 
lower, but nevertheless impressive, values measured in the isostructural 
family of frustrated triangular-lattice antiferromagnets,
Ca$_3$Co$_2$O$_6$ (7~T)~\cite{Hardy04}, and Ca$_3$CoMnO$_6$ (10~T)~\cite{Kim14,Jo09}, 
Sr$_3$CoIrO$_6$ ($\approx$~20~T)~\cite{Mikhailova12}, 
and Ca$_3$CoRhO$_6$ ($\approx$~30~T)~\cite{Niitaka01}.
Sr$_3$NiIrO$_6$ is initially antiferromagnetic after zero-field cooling and
its magnetic groundstate must evolve to produce a 
net magnetization \cite{LeFrancois14}.
The energy scale for the coercive field 
($\mu_0H_{\rm c} = 34 - 55$~T) is roughly of the 
same order of magnitude as the temperature of the 
onset of magnetic hysteresis ($13-28$~K), and long-range 
order (85~K), consistent with evolution of the magnetic order
driven by large magnetic fields. Indeed,
in compounds isostructural with Sr$_3$NiIrO$_6$, 
the microscopic order has been shown to change around the 
hysteresis loop using elastic neutron diffraction measurements \cite{Jo09,Fleck10}.
We can therefore infer that the magnetic hysteresis in Sr$_3$NiIrO$_6$
can be attributed to the evolution of a microscopic 
frustrated order with magnetic field, rather than domain effects
({\it c.f.} Ref.~\cite{Vilar11}).

As mentioned previously, electronic structure calculations suggest that
the magnetocrystalline anisotropy in Sr$_3$NiIrO$_6$ results 
from the configuration of overlapping orbitals in Ir$^{4+}$-O-Ni$^{2+}$ chain \cite{Zhang10,Sarkar10,Wu13}.
This picture is supported by the prevalence of relatively large coercive magnetic fields 
in the isostructural family members that have $4d$ or $5d$ ions 
(with relatively extended orbitals) on the octahedral site and $3d$ 
ions on the bipyramidal site: Ca$_3$CoRhO$_6$, Sr$_3$NiIrO$_6$ and 
Sr$_3$CoIrO$_6$~\cite{Flahaut03,Mikhailova12,Niitaka01}. 
The exceptional coercive magnetic field in the title compound serves as another 
example of notable physical effects resulting from the unusual
$5d$ orbital physics of Ir$^{4+}$.
\section{Conclusion}
In conclusion, we observed coercive magnetic fields of $34-55$~T for ${\bf H}\parallel {\bf c}$ 
in flux-grown single crystals and polycrystalline Sr$_3$NiIrO$_6$,
to our knowledge, a record high coercive magnetic field for any material. 
Sr$_3$NiIrO$_6$ shows signatures of a type of frustrated antiferromagnetic order 
in zero field \cite{LeFrancois14}, and the hysteresis loop is likely due to evolution of this microscopic order.
The high coercive magnetic field is consistent with the large 
magnetocrystalline anisotropy predicted in {\it ab-initio} calculations for this 
compound, due to the Ir$^{4+}$ $5d$ orbitals overlapping via intermediate oxygen with Ni$^{2+}$ $3d$ orbitals.

\section{Methods}
\subsection{Sample growth}
Polycrystalline Sr$_3$NiIrO$_6$ was 
prepared through solid-state reaction at $1300^o$C.
Single crystals of Sr$_3$NiIrO$_6$
were grown from either a
stoichiometric or 20~\% Ni-rich or 20~\% Ir-rich
(in molar ratio) composition using K$_2$CO$_3$ as flux.
The single crystals are hexagonal plates with
typical dimensions $2\times 2 \times 0.5$~mm$^3$.

\subsection{Magnetization measurements}
Quasistatic-field magnetization ($M(H)$) data were measured in 
a vibrating sample magnetometer in a superconducting 
magnet (PPMS-14, Quantum Design), or in a SQUID (MPMS-7, Quantum Design).
AC susceptibility data were measured in a 7~T 
AC SQUID and an ac susceptometer in a 14~T PPMS (Quantum Design). 

The pulsed-field magnetization experiments used a 1.5~mm bore, 1.5~mm long, 1500-turn
compensated-coil susceptometer, constructed from 50 gauge high-purity copper wire~\cite{Goddard08}.
When a sample is within the coil, the signal is $V \propto  ({\rm d}M/{\rm d}t)$,
where $t$ is the time. Numerical
integration is used to evaluate $M$. Samples were mounted within a 1.3~mm diameter
ampoule that can be moved in and out of the coil. Accurate values of $M$ are obtained
by subtracting empty-coil data from that measured under identical conditions with the sample
present. The susceptometer is calibrated by scaling low-field $M$ values to match those recorded with
a sample of known mass measured in a commercial SQUID or vibrating-sample magnetometer.

\begin{figure}
\centering
\includegraphics[width=0.44\textwidth]{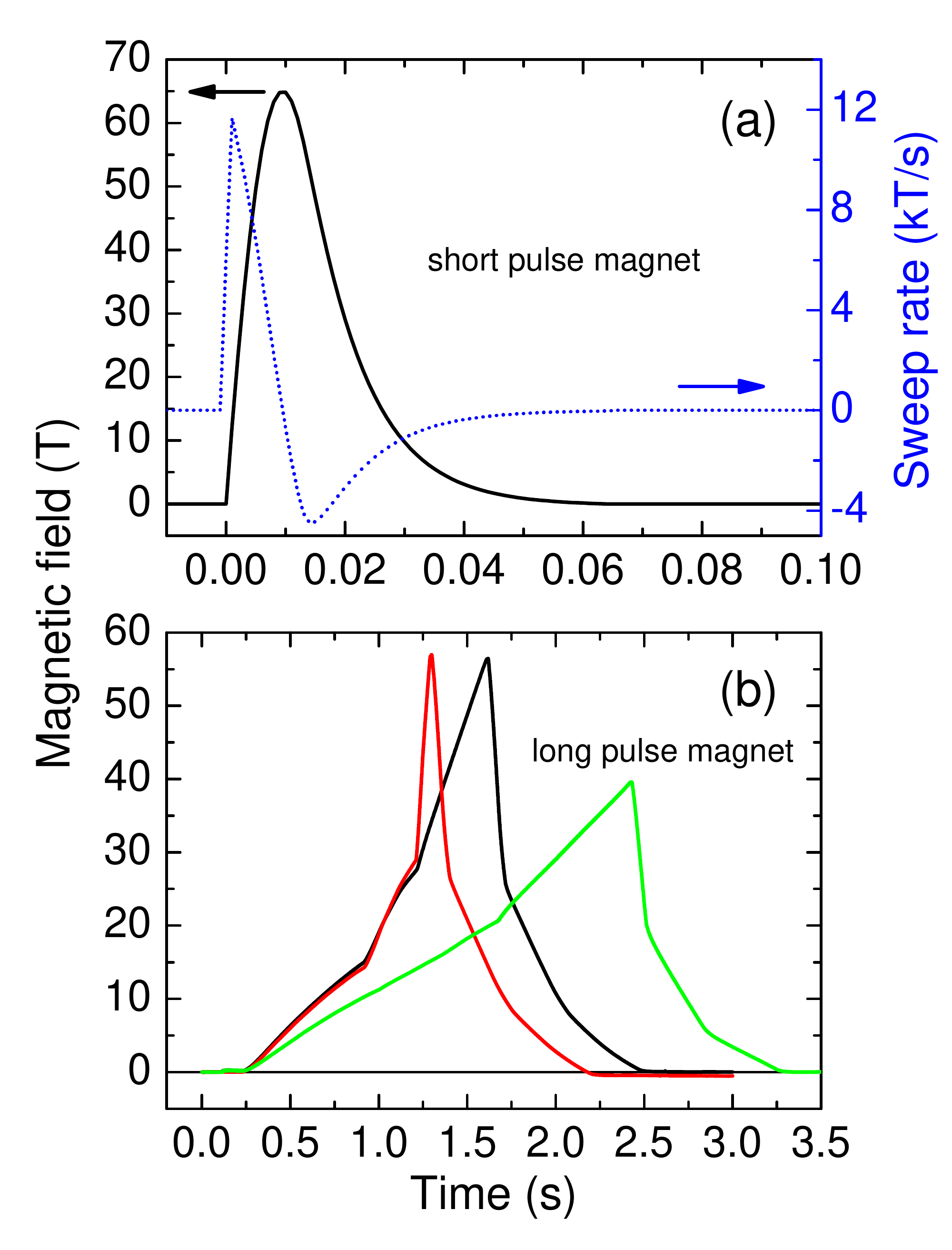}
\caption{{\bf Time-dependence of magnetic fields} for (a)~a~capacitor-bank-driven 
65~T short-pulse magnet and (b)~three examples of the controlled
sweep patterns possible with the~generator-driven 60~T Long Pulse Magnet.
(a)~$\mu_0 {\rm d}H/{\rm d} t$  for the short-pulse magnet is shown in blue (right axis).
In (b), the three stages in each of the field-sweep patterns
are due to three 
separate coils that are energized in sequence by the generator.}
\label{fieldbolg}
\end{figure}

Fields were provided by a 65~T short-pulse magnet energized by a 4~MJ capacitor bank,
or the generator-driven 60~T Long-Pulse Magnet at NHMFL Los
Alamos; the field versus time profiles for these two magnets
are shown in Fig.~\ref{fieldbolg}. The susceptometer was placed within a
$^3$He cryostat providing $T$s down to 0.4~K.
$\mu_0 H$ was measured by integrating the voltage induced in
a ten-turn coil calibrated by observing the de Haas-van Alphen oscillations of the belly orbits
of the copper coils of the susceptometer~\cite{Goddard08}.

In measuring hysteresis loops, the initial field sweep (up and down) is 
performed after zero-field cooling from room temperature.
Subsequent pulses are delivered after 40 to 60 minutes 
(the cooling time of the 
magnet in question) while maintaining constant sample temperature.

\section{Acknowledgments}
This work is funded by the U.S. Department of Energy, Basic Energy Sciences program ``Science at 100 Tesla".
The NHMFL Pulsed Field Facility is funded by the US National Science Foundation through Cooperative Grant No. DMR-1157490, the State of Florida, and the US Department of Energy.
Work in the UK is supported by the EPSRC.  PAG and JS would like to thank the University of 
Oxford for the provision of visiting fellowships.
The work at Rutgers was supported by the NSF under Grant number
NSF-DMREF-1233349, and the work at Postech
was supported by the Max Planck POSTECH/KOREA
Research Initiative Program (Grant number 2011-031558)
through NRF of Korea funded by MEST.

\section{Author Contributions}
J.S., J-W.K. and C.T. carried out the pulsed-field magnetization measurements, analyzed the data 
and prepared the figures in this paper.
C.T., P.A.G. and J.S. constructed the pulsed-field magnetometer.
E.D.M. assisted with pulsed-field experiments.
The SQUID and AC and DC magnetometry data in this paper
were measured by S.G., P.A.G. and A.H..
Samples were grown, characterized and prepared by X.L, Y.S.O. and S-W.C..
The text was written by J.S. and V.Z. with input from all authors. 
This study is part of a research program on multiferroics directed by V.Z.. 

\section{Additional information}
The authors declare no competing financial interests.


\end{document}